\documentclass[%
onecolumn,
nofootinbib,
 amsmath,amssymb,
 aps,
]{revtex4-2}

\usepackage{physics}
\usepackage[pdftex]{graphicx}
\usepackage{bm,float}
\usepackage{mathtools}
\usepackage{mathrsfs}
\usepackage{comment}
\usepackage{hyperref}
\hypersetup{
setpagesize=false,
bookmarksnumbered=true,
bookmarksopen=true,%
colorlinks=true,%
linkcolor=blue,
citecolor=blue,
}

\newcommand{\ep}{\epsilon}
\newcommand{\pa}{\partial}
\newcommand{\del}{\delta}

\usepackage{ulem}
\usepackage{xcolor}

\date{\today}

\begin{abstract} 
 In this work, we investigate the quantum entanglement properties of analog Hawking radiation produced by a moving mirror. Using two detector modes defined through window functions on a quantum field, we quantify the bipartite entanglement established between these modes. Our results reveal that the amount of entanglement accessible to the detectors increases when the mirror follows trajectories with non-monotonic, time-dependent acceleration, which are accompanied by the emission of negative energy flux. This indicates that the negative energy flux acts as a channel through which information can be returned. To substantiate this perspective, we examine how the recovery or reconstruction of the associated partner modes is related to the negative energy flux emitted by the mirror.
\end{abstract}   

\begin{document}
\title{Energy Flux as an Entanglement Current in Moving-Mirror Radiation}

\author{Yasusada Nambu}%
\email{yasusada.nambu@gmail.com}
\affiliation{%
Department of Physics, Graduate School of Science, Nagoya University, Nagoya
464-8602, Japan
}%
\affiliation{%
Kansai Institute for Photon Science (KPSI), National Institute for Quantum Science and Technology (QST), Kizugawa,
 Kyoto 619-0215, Japan
}%
\thanks{Visiting Collaborative Researcher at KPSI,QST}
\author{Riku Yoshimoto}
 \email{yoshimoto.riku.d1@s.mail.nagoya-u.ac.jp}

\affiliation{%
Department of Physics, Graduate School of Science, Nagoya University, Nagoya
464-8602, Japan
}%

\maketitle

\tableofcontents

\section{Introduction}
The moving-mirror setup has been investigated as a simplified model for Hawking radiation and  the evaporation process of black holes \cite{Birrell1984,Carlitz1987,Brout1995} due to its similarity to quantum radiation from black holes and its straightforward technical implementation. Evaporating scenarios are represented by the ray-tracing function, which defines the world line of the mirror. The quantum characteristics of the analog Hawking radiation produced by a moving mirror are fully determined by the behavior of the ray-tracing function.
 The information loss puzzle in evaporating black holes has been widely investigated in the framework of the moving mirror system \cite{Hotta2015,Tomitsuka2019,Chen2017a,Wald2019,Osawa2024a,agullo2025}. This issue is believed to be deeply tied to the fundamental nature of gravity and quantum physics. Once a black hole has completely evaporated, an initially pure quantum state appears to evolve into a thermal mixed state.  
In efforts to resolve this apparent violation of unitarity under physically reasonable assumptions, one attractive proposal is the vacuum fluctuation scenario \cite{Hotta2015,Wald2019,Osawa2024a,agullo2025}.
In this framework, vacuum fluctuations act as the purification partners of Hawking radiation \cite{Hotta2015,Tomitsuka2019} and are advantageous in that they do not pose any issues regarding energy costs \cite{Wald2019,Osawa2024a,agullo2025}.

 In the context of studying moving mirror systems from a quantum information perspective, detection of  entanglement from Hawking radiation has been  investigated~\cite{Cong2019,Tomonaga2024} based on the entanglement harvesting protocol using qubit detectors interacting with radiation from a moving mirror. In paper \cite{Cong2019}, the authors explore how entanglement is modified when influenced by the mirror’s emitted radiation. Focusing on mirrors that follow asymptotically future null trajectories and examining how the detectors’ distance from the mirror affects the outcome, they report the occurrence of entanglement sudden death induced by the radiation, as well as an enhancement of entanglement at specific detector separations. This enhancement is attributed to interference between the in-modes and out-modes.

 In this work, we study entanglement harvesting in the moving mirror setup and explore how information can be retrieved by examining the quantum correlations between two detector modes prepared at future null infinity. In particular, we concentrate on the emission of negative energy flux, which unavoidably arises for mirror trajectories that are asymptotically time-like, and aim to elucidate its relation to the unitary evolution of the system as discussed by Bianchi and Smerlak \cite{Bianchi2014a}.
 More concretely, we define two localized Gaussian modes (detector modes) A and B by specifying suitable profile functions of the scalar field. These detector modes are placed at future null infinity and serve as probes for the entanglement encoded in the radiation emitted by the mirror.  
We study three different classes of mirror trajectories and compare how the corresponding outgoing energy flux is related to the entanglement negativity of the detector modes.  
For mirror trajectories that generate episodes of negative energy flux, we find that during the intervals when this negative flux is emitted, the observed entanglement negativity exceeds the value associated with the Minkowski vacuum state.  The nature of the quantum correlations between detector modes A and B is strongly correlated with the dynamics of their partner particles. In the geometrical optics limit, the trajectories of these partner particles are, in fact, governed by the behavior of the ray-tracing function. 
We interpret and elucidate this phenomenon in terms of the recovery of partner particles and the conservation of information, i.e., the unitarity of the overall system.  

To further validate our understanding of entanglement behavior, we employ the partner formula \cite{Hotta2015,Tomitsuka2019} and explicitly derive the profile of the partner mode that purifies detector mode A. This construction supports our claim that the occurrence of a negative radiation flux genuinely indicates the retrieval of the partner particles.


The structure of this paper is as follows. In Sec. II, we introduce the moving mirror model and review the associated radiation emission. Section III is devoted to defining the local detector modes of the quantum field. In Sec. IV, we outline our entanglement harvesting setup in the context of the moving mirror. Numerical results are presented in Sec. V. Finally, Sec. VI contains our summary and conclusions. Throughout, we work in natural units with $c=\hbar=1$.

\section{Moving mirror model}
We consider a massless scalar field in the (1+1)-dimensional Minkowski
spacetime with a mirror boundary condition. The metric is
\begin{equation}
	ds^2=-dt^2+dx^2=-dudv,\quad u=t-x,\quad v=t+x.
\end{equation}
The trajectory of the mirror is described by $v = p(u)$, where $p(u)$ denotes the
ray-tracing function. In our analysis, we take $p(u)$ to be a non-decreasing function of $u$. For subsequent investigation into the behavior of entanglement, we examine the
three velocities and the proper acceleration of the mirror’s trajectory.  In
terms of the ray-tracing function, the three-velocity of the mirror trajectory
$x^\mu=(t,x(t))$ is
\begin{equation}
 \frac{dx}{dt}=\dot x=\frac{p'(u)-1}{p'(u)+1},
\label{eq:3-vel}
\end{equation}
and the proper acceleration is\footnote{The proper acceleration is defined as the
norm of the four-acceleration of the mirror $a=(a_\mu a^\mu)^{1/2},
~a^\mu=\frac{d^2x^\mu}{d\tau^2}$.}
\begin{equation}
 a=\frac{\ddot x}{(1-\dot x^2)^{3/2}}=\frac{p''(u)}{(p'(u))^{3/2}}.
\end{equation}

The scalar field $\phi(u,v)$   satisfies the Klein-Gordon equation
$\pa_u\pa_v\phi=0$ with the Dirichlet boundary condition at the mirror 
$\phi(u,p(u))=0$, and the general solution is given by
\begin{equation}
	\phi(u,v)=\phi^\text{in}(v)+\phi^\text{out}(u),\quad \phi^\text{out}(u)=-\phi^\text{in}(p(u)),
	\label{eq:in-out}
\end{equation}
where $\phi^\text{in}(v)$ and $\phi^\text{out}(u)$ are arbitrary functions of $v$ and $u$, respectively. 
Assuming the plane-wave as the in-mode and out-mode waves, the normal mode  is expressed as
\begin{subequations}
\begin{align}
	&\phi^\text{in}_\omega(v)=\frac{1}{\sqrt{4\pi\omega}}e^{-i\omega v},\quad\omega>0, \quad \text{on the past null infinity~}\mathscr{I}^{-}\label{eq:mode} \\
    &\phi^\text{out}_\omega(u)=\frac{1}{\sqrt{4\pi\omega}}e^{-i\omega u},\quad\omega>0, \quad \text{on the future null infinity~}\mathscr{I}^{+}
\end{align}
\end{subequations}
With the mirror boundary condition, the scalar field operator on $\mathscr{I}^+$ is expressed as
\begin{subequations}
\begin{align}
  \hat\phi(u)&=-\int_0^\infty d\omega\left[\hat a_\omega \,\phi^\text{in}_\omega(p(u))+\hat
    a_\omega^{\dag}\,\phi_\omega^{\text{in}*}(p(u))\right],\\
&=\int_0^\infty d\omega\left[\hat b_\omega \,\phi^\text{out}_\omega(u)+\hat
    b_\omega^{\dag}\,\phi_\omega^{\text{out}*}(u)\right],
\end{align}
\end{subequations}
where $\hat a_\omega$ is the annihilation operator of the in-vacuum state $\hat a_\omega|0_\text{in}\rangle=0$, and $\hat b_\omega$ is the annihilation 
operator of the out-vacuum state $\hat b_\omega|0_\text{out}\rangle=0$.  We assume the in-vacuum state of the field operator to evaluate radiation from the mirror. If we are interested in detecting analog Hawking
radiation emanating from the mirror on the future null infinity
$\mathscr{I}^+$, it is sufficient to consider only the right-moving modes of
the scalar field. The field operator  $\hat\phi$ and its gauge-invariant field operator $\hat\Pi:=\pa_u\hat\phi$ can be expressed as
\begin{subequations}
\begin{align}
 &\hat\phi(u)=-\int_0^\infty\frac{d\omega}{\sqrt{4\pi\omega}}\left[\hat a_\omega
\,e^{-i\omega p(u)}+\hat a_\omega{}^{\!\!\!\!\dag}\,\, e^{i\omega p(u)}\right], \\
&\hat\Pi(u)=ip'(u)\int_0^\infty d\omega\sqrt{\frac{\omega}{4\pi}}\left[\hat
a_\omega\, e^{-i\omega p(u)}-\hat a_\omega{}^{\!\!\!\!\dag}\,\, e^{i\omega p(u)}\right],\\
&[\hat\phi(u_1),\hat\Pi(u_2)]=\frac{i}{2}\,\del(u_1-u_2),\quad
	[\hat\Pi(u_1),\hat\Pi(u_2)]=\frac{i}{2}\,\del'(u_1-u_2).
\end{align}
\end{subequations}

For these  operators, correlation functions with respect to the in-vacuum state prepared on $\mathscr{I}^-$ are
\begin{subequations}
\begin{align}
	&\expval{\hat\phi(u_1)\hat\phi(u_2)}_\text{in}=-\frac{1}{4\pi}\ln(p(u_1)-p(u_2
)-i(\Delta p)_{12}),\\
	&\expval{\hat\Pi(u_1)\hat\Pi(u_2)}_\text{in}=-\frac{1}{4\pi}\frac{p_1'\,p_2'}{(p
_1-p_2-i(\Delta p)_{12})^2},\\
	&\expval{\{\hat\Pi(u_1),\hat\Pi(u_2)\}}_\text{in}=\frac{p_1'p_2'}{2\pi}\int_0
^\infty d\omega\,\omega\cos(\omega(p_1-p_2))\,e^{-(\Delta
p)_{12}\omega}=:D_\text{in}(u_1,u_2),
\end{align}
\end{subequations}
where  a UV cutoff $(\Delta p)_{12}>0$ is introduced by
\begin{equation}
  (\Delta p)_{12}:=p'\left(\frac{u_1+u_2}{2}\right)\ep,\quad \ep>0,
\end{equation}
with a positive small parameter $\ep$ that defines the UV cutoff length of the
in-vacuum state. Thus $(\Delta p)_{12}$ is the UV cutoff length of the out-vacuum state.

After removing the divergent part associated with Minkowski vacuum fluctuations, the regularized energy flux measured at future null infinity can be written as
\cite{Birrell1984}
\begin{align}
F(u):&=\langle 0_\text{in}|\hat T_{uu}(u)|0_\text{in}\rangle=\lim_{\del u\rightarrow 0}\left\{\expval{\hat\Pi(u)\hat\Pi(u+\del u)}_\text{in}-\expval{\hat\Pi(u)\hat\Pi(u+\del u)}_\text{out}\right\} \notag \\
&=-\frac{1}{24\pi}
\left[\frac{p'''(u)}{p'(u)}-\frac{3}{2}\left(\frac{p''(u)}{p'(u)}\right)^2\right] \label{eq:Tu} \\
%
%
&=-\frac{1}{24\pi}(p'(u))^{1/2}a'(u).
\notag
\end{align}
Thus, the sign of $F(u)$ is determined by the sign of $a'(u)$: the rate of change
of the proper acceleration. For a mirror’s trajectory with a  period denoted as $a'(u)>0$, the function $F(u)$ becomes negative within that regime.

\section{Local modes in quantum field}
We examine the entanglement properties of the radiation emitted by the mirror. To this end, we define two detector modes, A and B, designed to measure the entanglement present in the radiation. In order to obtain the local detector modes A and B from the quantum field, we introduce the following complex mode profile functions, constructed as superpositions of the Minkowski mode functions:
\begin{equation}
	\xi_j(u)=\int_{-\infty}^\infty d\omega\, \widetilde \xi_j(\omega) \frac{e^{-i\omega
u}}{\sqrt{4\pi|\omega|}},\quad j=A,B,
\end{equation}
where $\widetilde\xi_j(\omega)$ is the weighting function that defines the shape of the local mode. The profile function is normalized as
\begin{equation}
	(\xi_{j_1}(u),\xi_{j_2}(u))_\text{KG}=i\int_{-\infty}^\infty
du\,\xi_{j_1}^*(u)\overset{\leftrightarrow}{\partial}_u\xi_{j_2}(u)=2i\int_{-\infty}
^\infty du\, \xi_{j_1}^*(u)\xi_{j_2}'(u)\equiv \del_{j_1j_2},\quad j_1,j_2=A,B.
\end{equation}
Consequently, the weighting function satisfies the following normalization
condition:
\begin{equation}
	\int_{-\infty}^\infty d\omega\,
\widetilde{\xi}_{j_1}^*(\omega)\widetilde\xi_{j_2}(\omega)\,\mathrm{sgn}(\omega)=\del_{j_1j_2}. 
\end{equation}
For the mode profile $\xi(u)$, the corresponding annihilation and creation operators are defined as follows
\begin{equation}
	\hat a_j=(\xi_j,\hat\phi)_\text{KG}=2i\int_{-\infty}^\infty du\,
\xi_j^*(u)\hat\Pi(u),\quad
	\hat a_j{}^{\!\!\dag}=(\xi^*_j,\hat\phi)_\text{KG}=-2i\int_{-\infty}^\infty
du\, \xi_j(u)\hat\Pi(u).
\end{equation}
The commutator between these operators satisfies
\begin{equation}
	[\hat a_{j_1},\hat a_{j_2}{}^{\!\!\dag}]=2i\int_{-\infty}^\infty
du\,\xi_{j_1}^*(u)\xi_{j_2}'(u)=\del_{j_1j_2}\quad [\hat a_{j_1},\hat a_{j_2}]=0,\quad j_1,j_2=A,B.
\end{equation}
Canonical operators related to $\hat a_j$ are defined by the following
relations:
\begin{equation}
	\hat a_A=\frac{\hat Q_A+i\hat P_A}{\sqrt{2}},\quad\hat a_B=\frac{\hat
Q_B+i\hat P_B}{\sqrt{2}}.
\end{equation}
Using mode profile functions $Q_j(u),P_j(u), j=A,B$, the canonical operators
are represented as
\begin{align}
&\hat Q_j=\int_{-\infty}^\infty du\, Q_j(u)\,\hat\Pi(u),\quad \hat
P_j=\int_{-\infty}^\infty du\,
  P_j(u)\,\hat\Pi(u),
\end{align}
with the normalization condition imposed by the canonical commutation relation:
\begin{align}
	&\int_{-\infty}^\infty du\, Q_{j_1}(u)P'_{j_2}(u)=-\int_{-\infty}^\infty du\, Q'_{j_1}(u)P_{j_2}(u) =2\del_{j_1j_2},\quad j_1,j_2=A,B,
	\label{eq:PQnorm}
\end{align}
where we assume that the surface terms of the integral vanish.
The profiles $\xi_j(u)$ and $P_j(u),Q_j(u)$ are related as
\begin{equation}
	Q_j(u)=i\sqrt{2}\,(\xi_j^*(u)-\xi_j(u))=2\sqrt{2}\,\mathrm{Im}(\xi_j),\quad
P_j(u)=\sqrt{2}\,(\xi_j^*(u)+\xi_j(u))=2\sqrt{2}\,\mathrm{Re}(\xi_j).
\end{equation}
Covariances of the local canonical operators are
\begin{subequations}
\begin{align}
	&\expval{\{\hat Q_{j_1},\hat Q_{j_2}\}}_\text{in}=\int_{-\infty}^{\infty}
du_1du_2\,Q_{j_1}(u_1)Q_{j_2}(u_2)\expval{\{\hat\Pi_1,\hat\Pi_2\}}_\text{in}, \\
	&\expval{\{\hat Q_{j_1},\hat P_{j_2}\}}_\text{in}=\int_{-\infty}^\infty
du_1du_2\,Q_{j_1}(u_1)P_{j_2}(u_2)\expval{\{\hat\Pi_1,\hat\Pi_2\}}_\text{in},\\
	&\expval{\{\hat P_{j_1},\hat P_{j_2}\}}_\text{in}=\int_{-\infty}^\infty
du_1du_2\,P_{j_1}(u_1)P_{j_2}(u_2)\expval{\{\hat\Pi_1,\hat\Pi_2\}}_\text{in}.
\end{align}
\end{subequations}
The covariance matrix of the bipartite state AB is
\begin{equation}
	V_{AB}=\begin{bmatrix}
		a_1 &a_3&c_1&c_3 \\
		a_3&a_2&c_4&c_2\\
		c_1&c_4&b_1&b_3\\
		c_3&c_2&b_3&b_2
	\end{bmatrix},
\end{equation}
where its components are given by
\begin{subequations}
\begin{align}
	&c_1=\expval{\{\hat Q_A,\hat Q_B\}}_\text{in},\quad c_2=\expval{\{\hat P_A,\hat
P_B\}}_\text{in},\\
	&c_3=\expval{\{\hat Q_A,\hat P_B\}}_\text{in},\quad c_4=\expval{\{\hat P_A,\hat
Q_B\}}_\text{in},\\
	&a_1=2\expval{\hat Q_A^2}_\text{in},\quad a_2=2\expval{\hat P_A^2}_\text{in},\quad
a_3=\expval{\{\hat Q_A,\hat P_A\}}_\text{in},\\
	&b_1=2\expval{\hat Q_B^2}_\text{in},\quad b_2=2\expval{\hat P_B^2}_\text{in},\quad
b_3=\expval{\{\hat Q_B,\hat P_B\}}_\text{in}.
\end{align}
\end{subequations}
The entanglement entropy of local mode A is obtained by
\begin{subequations}
\begin{align}
&S_A=\left(\frac{\nu+1}{2}\right)\log_2\left(\frac{\nu+1}{2}\right)-\left(\frac{\nu-1}{2}\right)\log_2\left(\frac{\nu-1}{2}\right), \label{eq:SA-nu}\\
&\nu=\sqrt{a_1a_2-a_3^2}, \label{eq:nuA}
\end{align}
\end{subequations}
where $\nu$ is the symplectic eigenvalue of local mode A.
The separability of the bipartite system AB can be evaluated in terms of the
logarithmic negativity \cite{Vidal2002a,Plenio2005}:
\begin{equation}
  E_N:=\mathrm{max}[-\log_2\tilde\nu,0],
\end{equation}
where $\tilde\nu$ denotes the symplectic eigenvalue of the partially transposed
 covariance  matrix with $\tilde\nu<1$. For $E_N=0$, the bipartite state AB is separable, and for $E_N>0$, the bipartite state AB is
entangled; the logarithmic negativity represents the amount of entanglement
between A and B.


\section{Entanglement harvesting in  moving mirror model}

\subsection{Setup of detector modes A and B}
\begin{figure}[bht]
\centering
\includegraphics[width=0.6\linewidth]{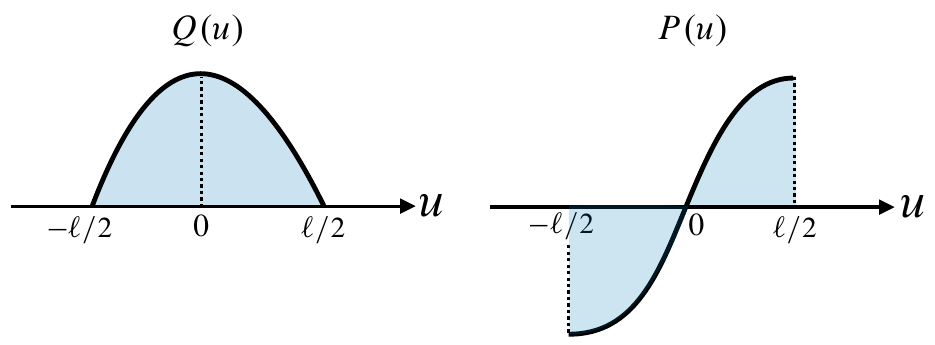}
\caption{Profile functions $Q(u)$ and $P(u)$. They have compact support
$u\in[-\ell/2,\ell/2]$.}
\label{fig:PQ}
\end{figure}

\begin{figure}[bht]
\centering
\includegraphics[width=0.9\linewidth]{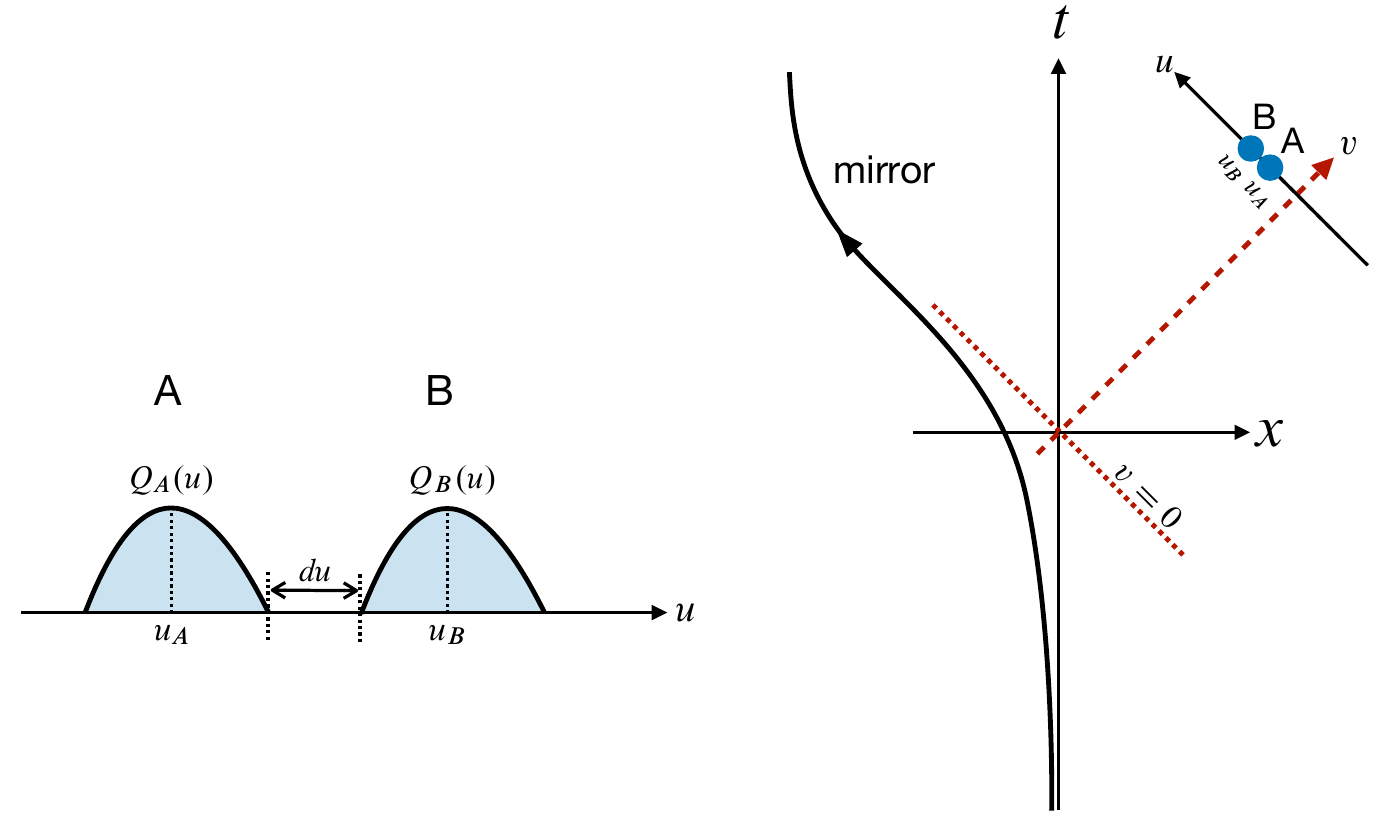}	
\caption{Setup for detecting entanglement of the analog Hawking radiation from a moving
mirror. Left panel: We show the configuration of profile functions $Q_A(u)$ and $Q_B(u)$. The bipartite local mode AB captures the entanglement of the 
radiation. Right panel: we observe the dependence of the harvested entanglement at the location of the adjacent two local modes A and B ($du=0$).}
\label{fig:setup1}
\end{figure}

We assume the following compact weighting function of local modes (Fig.
\ref{fig:PQ}):
\begin{align}
	&Q(u)=\frac{2}{\sqrt{\pi}}\cos\left(\frac{\pi u}{\ell}\right), \quad
P(u)=\frac{2}{\sqrt{\pi}}\sin\left(\frac{\pi u}{\ell}\right),\quad u\in
[-\ell/2,\ell/2].
\label{eq:profile}
\end{align}
\noindent
These profile functions satisfy the normalization \eqref{eq:PQnorm}. For the
purpose of entanglement harvesting from the quantum field, we prepare two local
modes A and B (detector modes) as shown in Fig. \ref{fig:setup1}:
\begin{equation}
Q_A(u)=Q(u-u_A),\quad P_A(u)=P(u-u_A),\quad Q_B(u)=Q(u-u_B),\quad
P_B(u)=P(u-u_B).
\end{equation}
The components of the covariance matrix of the bipartite state AB are
\begin{subequations}
\begin{align}
	&a_1=\int_{-\ell/2}^{\ell/2}du_1du_2Q(u_1)Q(u_2)D_\text{in}(u_1+u_A,u_2+u_A), \\
	&a_2=\int_{-\ell/2}^{\ell/2}du_1du_2P(u_1)P(u_2)D_\text{in}(u_1+u_A,u_2+u_A), \\
	&a_3=\int_{-\ell/2}^{\ell/2}du_1du_2Q(u_1)P(u_2)D_\text{in}(u_1+u_A,u_2+u_A), \\
	&b_j=a_j(u_A\rightarrow u_B),\quad j=1,2,3, \\
	&c_1=\int_{-\ell/2}^{\ell/2}du_1du_2Q(u_1)Q(u_2)D_\text{in}(u_1+u_A,u_2+u_B), \\
	&c_2=\int_{-\ell/2}^{\ell/2}du_1du_2P(u_1)P(u_2)D_\text{in}(u_1+u_A,u_2+u_B), \\
	&c_3=\int_{-\ell/2}^{\ell/2}du_1du_2Q(u_1)P(u_2)D_\text{in}(u_1+u_A,u_2+u_B).\\
	&c_4=\int_{-\ell/2}^{\ell/2}du_1du_2P(u_1)Q(u_2)D_\text{in}(u_1+u_A,u_2+u_B).
\end{align}
\end{subequations}

Figure \ref{fig:setup1} shows our setup for entanglement harvesting with the
moving mirror system. Our focus is on the dependence of the harvestable
entanglement on the location  $u_A$ of the detector modes.

\section{Behavior of entanglement negativity}
\subsection{Mirror models}
\begin{figure}[bht]
\centering
\includegraphics[width=1\linewidth]{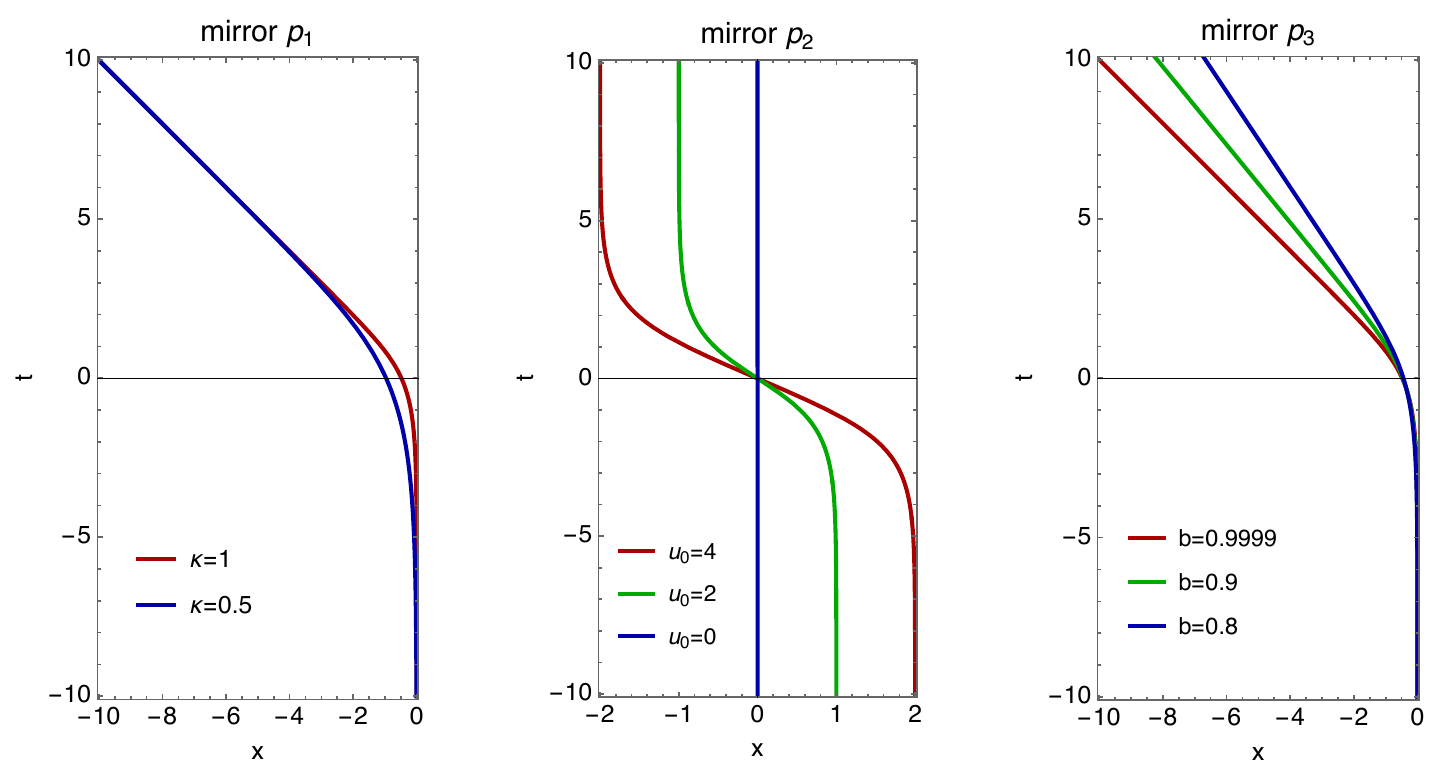}	
\caption{Three classes of trajectories associated with the ray-tracing functions $p_1$, $p_2$, and $p_3$. For $p_2$, we display trajectories for the case $\kappa = 1$.}
\label{fig:trj}
\end{figure}

We examine three different mirror trajectories (Fig.~\ref{fig:trj}).
The first trajectory is the mirror with the ray-tracing function
\begin{equation}
p_1(u)=-\frac{1}{\kappa} \ln(1+e^{-\kappa u}). 
\end{equation}
In the asymptotic past $u=-\infty$, the mirror is static, and then the function approaches  $p_1 \sim -e^{-\kappa u}/\kappa$ as $u\rightarrow +\infty$. In this limit, the mirror’s velocity approaches the speed of light, 
effectively becoming an eternal mirror. Consequently, the mirror emits a
constant energy flux in the asymptotic future (see the left panel of Fig.~\ref{fig:trj1}), as demonstrated by Carlitz and
Willey \cite{Carlitz1987}:
\begin{equation}
  F(u)=\frac{\kappa^2}{48\pi}.
\end{equation}

The second trajectory is determined by the ray-tracing function
\begin{equation}
p_2(u)=\frac{1}{\kappa}
\ln\left(\frac{1+e^{\kappa(u-u_0)}}{1+e^{-\kappa(u+u_0)}}\right). 
\end{equation}
Starting from a static state, the mirror undergoes acceleration and
subsequent deceleration to come to a complete halt (kink mirror). Using \eqref{eq:3-vel}, the maximum three velocity  at $u=0\, (t=0)$ is
\begin{equation}
 \dot x=-\frac{e^{\kappa u_0}-1}{e^{\kappa u_0}+3}.
\end{equation}

The third trajectory is determined by the ray-tracing function
\begin{equation}
p_3(u)=(1-b)u-\ln(1+e^{-bu}),\quad b\leq 1.
\end{equation}
Initially, the mirror possesses zero velocity. Subsequently, it undergoes
acceleration, culminating in inertial motion characterized by a non-zero
velocity.  Using \eqref{eq:3-vel}, the value of the final three-velocity of the mirror is determined by the parameter $b$:
\begin{equation}
  \dot x\sim -\frac{b}{2-b}\quad \text{as}\quad u\rightarrow\infty.
\end{equation}
$b=1$ reduces to the ray-tracing function $p_1(u)$ with $\kappa=1$.

\begin{figure}[bht]
\centering
\includegraphics[width=1\linewidth]{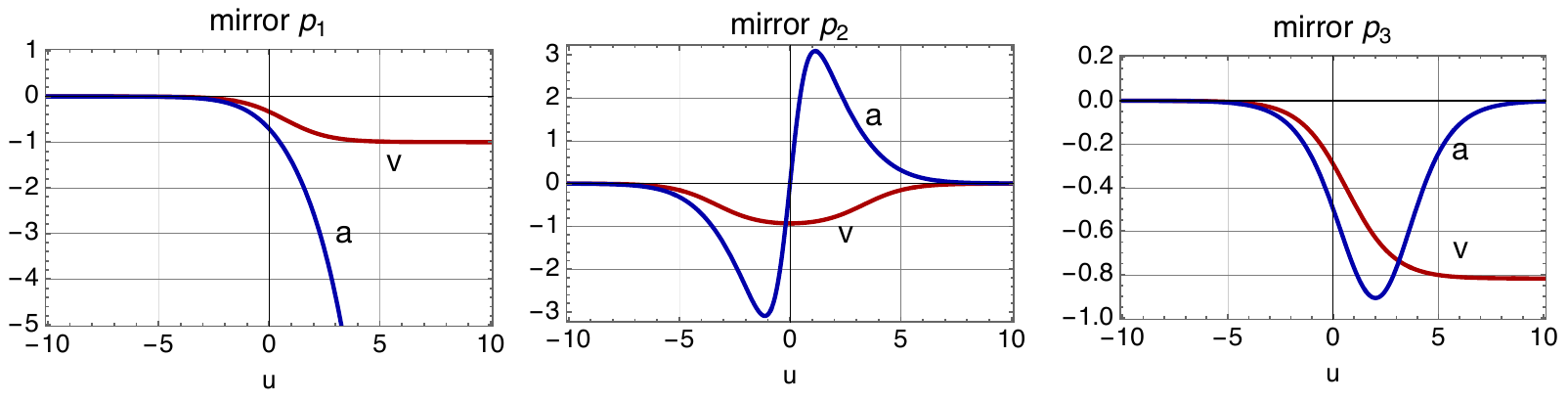}	
\caption{The three velocities $v$ and the corresponding proper accelerations $a$ for each mirror trajectory are shown. The parameters are chosen as $\kappa=1$ for $p_1$, $u_0=4$ for $p_2$, and $b=0.9$ for $p_3$.}
\label{fig:va}
\end{figure}

Figure \ref{fig:va} depicts the evolution of the three velocities and the proper accelerations of the mirrors. The acceleration of mirror $p_1$ decreases monotonically, while the three velocities asymptotically approach unity (the speed of light). Mirror $p_2$ first accelerates and then decelerates; as a result, its proper acceleration changes sign and exhibits two extremal points. Mirror $p_3$ undergoes a phase of acceleration followed by inertial motion, and thus its proper acceleration features a single extremum.
\noindent
The pattern of the proper acceleration is connected to the emitted flux \eqref{eq:Tu}. For mirrors characterized by the ray-tracing functions $p_2$ and $p_3$, there is a positive interval $a'$ over which the energy flux takes on negative values (see the left panels of Figs. \ref{fig:trj2} and \ref{fig:trj3}).

We compute the entanglement negativity of the bipartite detector system AB numerically. We adopt the width of the
detector mode $\ell=1$ and the value of the UV cutoff parameter $\ep=0.005$ in the following numerical computations.

\subsubsection{Eternal mirror $p_1$}
\begin{figure}[bht]
\centering
\includegraphics[width=0.85\linewidth]{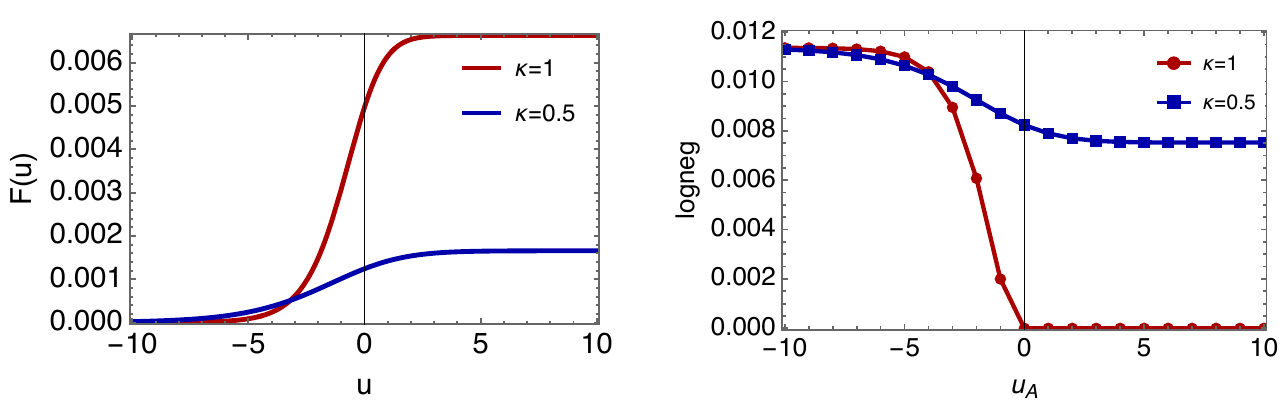}	
\caption{ Behavior of the energy flux and the negativity for the mirror trajectory $p_1$. Left panel: energy flux. Right panel: negativity as a function of $u_A$. The markers represent data points obtained from numerical computations.}
\label{fig:trj1}
\end{figure}

Figure \ref{fig:trj1} shows the $u$-dependence of the energy flux (left panel) and
the negativity of the detector state AB (right panel). For this trajectory, the energy flux is always positive, corresponding to  $a'<0$ (see Eq.\eqref{eq:Tu}). Negativity decreases
monotonically, and entanglement death is observed for the $\kappa=1$ case. We can interpret this behavior as the flux of Hawking radiation disrupting the coherence of the bipartite state AB, leading to the breaking of the quantum correlation between A and B.  Similar behavior of entanglement has already been discussed in the analog Hawking radiation of the quantum Hall system with an expanding edge
\cite{Nambu2023a,Yoshimoto2025}.

\subsubsection{Kink mirror $p_2$}
\begin{figure}[bht]
\centering
\includegraphics[width=0.9\linewidth]{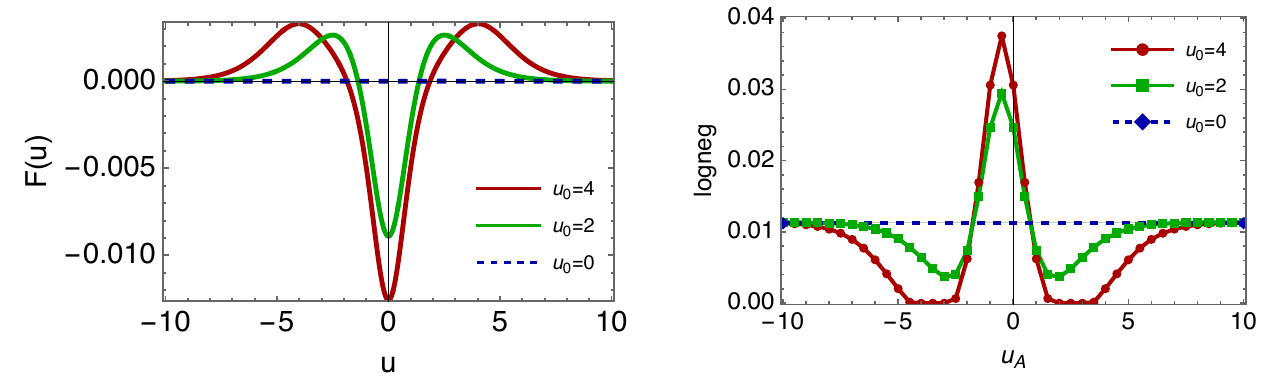}	
\caption{Behavior of the energy flux and the negativity for the mirror trajectory $p_2$ with $\kappa=1$. Left panel: energy flux. Right panel: negativity as a function of $u_A$. The parameter $u_0=0$ represents a static mirror trajectory, which does not radiate but maintains a constant, non-zero entanglement of the vacuum fluctuations. Markers show data points obtained from numerical computations.}
\label{fig:trj2}
\end{figure}

Figure \ref{fig:trj2} presents the $u$-dependence of the energy flux and the negativity of the detector state AB. Along this trajectory, the energy flux becomes negative when the time derivative of the acceleration, $a'$, changes sign from negative to positive. The evolution of the negativity follows the pattern of the emitted energy flux quite closely. It first decreases, then increases beyond its initial value, and reaches its maximum at the point where the negative energy flux is most pronounced. During the interval in which negative energy flux is emitted, characterized by $a' > 0$, the negativity exceeds its value in the vacuum limit ($u\rightarrow\pm\infty$). In the next section, we will examine this behavior of the negativity in more detail, based on the first law of entanglement and  the entanglement monogamy property.

\subsubsection{Asymptotically inertial mirror $p_3$}
\begin{figure}[bht]
\centering
\includegraphics[width=0.9\linewidth]{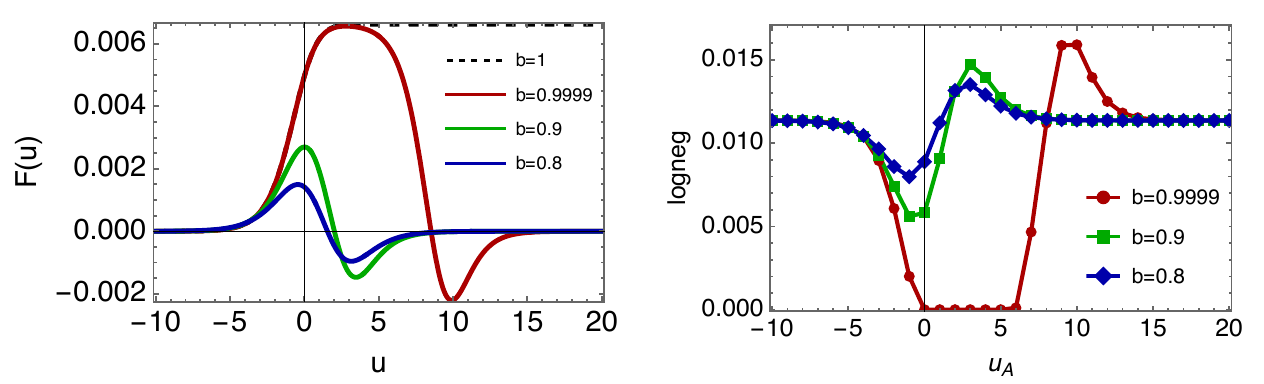}	
\caption{Behavior of the energy flux and the negativity for the mirror trajectory $p_3$. The case $b=1$ corresponds to the eternal mirror trajectory $p_1$ with $\kappa=1$. Left panel: energy flux. Right panel: negativity as a function of $u_A$. The markers indicate the data points obtained from numerical computations.}
\label{fig:trj3}
\end{figure}

Figure \ref{fig:trj3} depicts how the energy flux and the negativity of the detector pair AB vary with $u$. The energy flux turns negative when the acceleration’s magnitude ceases to grow and the motion becomes inertial. The negativity initially decreases but later grows and eventually exceeds its initial value, which equals the negativity of the Minkowski vacuum, at the point where the acceleration stops decreasing.

\subsection{Discussion}
We observed that a positive energy flux diminishes the amount of negativity, whereas a negative energy flux increases it. We justify the behavior of negativity by relating it to the properties of entanglement and by interpreting the energy flux as a measure of entanglement flow.

\subsubsection{Behavior of entanglement entropy and first law of entanglement in moving mirror systems}

Following the paper by Bianchi and Smerlak \cite{Bianchi2014a}, the regularized  entropy of the
interval $[u_0,u]$ in $\mathscr{I}^+$ is given by the  covariant regularization
scheme with broadening of the interval. The formula is
\begin{equation}
 S(u_0,u)=\frac{c}{12}\log\left[\frac{(p(u)-p(u_0))^2}{p'(u)p'(u_0)(u-u_0
)^2}\right],
\label{eq:S}
\end{equation}
where $c$ is the central charge of the conformal field. In the paper \cite{Bianchi2014a}, the authors take the limit $u_0\rightarrow -\infty$ to obtain the relation between the entanglement entropy of the interval $[-\infty,u]$:
\begin{equation}
S(u)=-\frac{c}{12}\log p'(u)=\frac{c}{12}\int^u du' \kappa(u'),
\end{equation}
where $\kappa(u):=-p''(u)/p'(u)$ is the ``peeling factor" \cite{Barcelo2011b}: for the eternal mirror $p_1$, $\kappa(u)$ asymptotically tends to a constant whose value matches the surface gravity of the Rindler horizon. For arbitrary mirror trajectories, its value as a function of time represents the extent of particle production caused by the mirror’s accelerated motion. This entropy is demonstrated to be related to the energy flux through
\begin{equation}
 F(u)=\frac{1}{2\pi}\left(\frac{6}{c}(S'(u))^2+S''(u)\right).
\end{equation}
From this formula, they inferred that the emission of negative energy flux is unavoidable if the unitary evolution of the system is to be maintained: $\lim_{u\rightarrow\infty}S(u)=0$. In our analysis, we do not take the $u_0\rightarrow -\infty$ limit in \eqref{eq:S}. Instead, we focus on the case where the interval is small compared with the characteristic variation scale of the mirror trajectory, and we examine how the local entropy of this small region A is related to the correlations in the mirror radiation as measured by the bipartite system AB.  We apply the formula \eqref{eq:S} 
to estimate the entropy of the detector mode A, which has support $[u_A-\ell/2, u_A+\ell/2]$ in our numerical calculation. Thus, the entanglement entropy of mode A, which is defined as the deviation from the vacuum entanglement, is given by
\begin{align}
S_A&=\frac{1}{12}\log\left[\frac{(p(u_A+\ell/2)-p(u_A-\ell/2))^2}{p'(u_A+\ell/2
)p'(u_A-\ell/2)\ell^2}\right] \label{eq:sA} \\
&\approx \frac{1}{12}\log\left[1-\frac{\ell^2}{6}\left(\frac{p'''}{p'}
-\frac{3}{2}\left(\frac{p''}{p'}\right)^2\right)+O(\ell^4)\right] \notag \\
&\approx \frac{\pi}{3} F(u_A)\ell^2,
\label{eq:1stlaw}
\end{align}
where we assumed that the size of the region $\ell$ is small, as stated in the second line. 
This approximation is justified if  $\ell$ is  sufficiently smaller than the changing scale determined by the ray-tracing function: $\ell\ll |p/p'|$. We notice that
the equation \eqref{eq:1stlaw} is a realization of  the first law of entanglement entropy 
\begin{equation}
\del S = \del\expval{H_A},
\label{eq:1stlaw}
\end{equation}
where $H_A$ is the modular Hamiltonian \cite{Blanco2013,Faulkner2014}. This relation is the analog of the first law of thermodynamics and encodes energy conservation for small perturbations. In our setup, the modular Hamiltonian is identified with the energy–momentum tensor of the scalar field that generates the Lorentz boost. Consequently, for a sufficiently small interval, the variation of the entanglement entropy follows the same pattern as the energy flux through that region: when a positive energy flux traverses region A, the entropy $S_A$ increases, while a negative energy flux through region A leads to a decrease in $S_A$. The first law of entanglement \eqref{eq:1stlaw} in the mirror system therefore implies that the energy flux of the radiation acts as a local “entanglement current’’.

It is possible to obtain this relation for Gaussian systems without referring to the specific details of the system.
 For this purpose, we consider the variation of the entanglement entropy. 
We denote the variation of the density matrix as follows,
\begin{equation}
    \rho=\rho_0 + \delta \rho,
\end{equation}
and the variation of the entanglement entropy $S=-\Tr [\rho\log \rho]$ is written as
\begin{equation}
    \begin{split}
        \delta S:&=S(\rho)-S(\rho_0)\\
        &=-\Tr \left[\delta\rho\log\rho_0+\rho_0\delta(\log\rho)\right]+\mathcal{O}((\delta \rho)^2)\\
        &=-\Tr\left[\delta\rho\log \rho_0\right].
    \end{split}
\end{equation}
In the last line, we used $\Tr \left[\rho_0\,\delta\!\log \rho\right]=0$. The derivation of this equation is presented in Appendix \ref{sec:appendixB}.

By introducing the modular Hamiltonian $H_A=-\log \rho_0$, $\delta S$ can be interpreted as the variation of the expectation value of the modular Hamiltonian:
\begin{equation}\label{delta S}
    \delta S=\Tr [H_A\delta \rho]=\Tr[H_A(\rho-\rho_0)]=\expval{H_A}-\expval{H_A}_0:=\delta\langle H_A \rangle.
\end{equation}
To obtain the first law like relation Eq.~\eqref{eq:1stlaw}, we need the explicit form of the modular Hamiltonian. We assume a single mode Gaussian state with zero-mean (Indeed, the local modes that we consider in this paper satisfy these conditions) and let us introduce the vector notation of the canonical variables
\begin{equation}
    \hat{\mathbf R}=(\hat Q,\hat P)^T,
    \qquad [\hat R_i,\hat R_j]=i\Omega_{ij},
    \qquad
    \Omega=\begin{pmatrix}0&1\\-1&0\end{pmatrix}.
\end{equation}
The covariance matrix of this Gaussian state is
\begin{equation}
    V_{ij}=\langle\{\hat R_i,\hat R_j\}\rangle.
\end{equation}
For Gaussian states, it is possible to write down the form of the density matrix using the techniques in \cite{Banchi2015}
\begin{equation}
    \rho_0=\frac{1}{Z}\exp\left[-\frac12\hat{\mathbf R}^{T}G\hat{\mathbf R}\right],\quad 
    G=\epsilon(\nu)\, \nu\, V^{-1} ,
\end{equation}
where $Z$ is a normalization constant independent of the variation, $\nu$ is the symplectic eigenvalue of this single-mode Gaussian state, and $\epsilon(\nu)$ is  defined as $\epsilon(\nu):=\log \frac{\nu +1}{\nu-1}$. Using these equations, the modular Hamiltonian is given by
\begin{equation}
    H_A=-\log\rho=\frac12\hat{\mathbf R}^{T}G\hat{\mathbf R}+\log Z.
\end{equation}
Using this Hamiltonian, the variation of the entropy \eqref{delta S} becomes
\begin{equation}
    \begin{split}
        \delta S&=\delta\langle H_A\rangle
        =\frac{1}{2}\Tr \left[\delta \rho\,\hat{\mathbf{R}}^TG\hat{\mathbf{R}}\right]
        =\frac{1}{2}\sum_{ij}G_{ij}\Tr \left[\delta\rho\, \hat{R}_i\hat{R}_j\right]\\
        &=\frac{1}{4}\sum_{ij}G_{ij}\delta V_{ij}
        =\frac{\epsilon(\nu)\, \nu }{4}\Tr [V^{-1}\delta V]=\frac{\epsilon(\nu)}{2}\,\delta\nu,
    \end{split}
\end{equation}
where $\delta\nu$ is the variation of the symplectic eigenvalue and we used the symmetric property of the matrix  $G_{ij}$.
To apply this equation for a local Gaussian mode in moving mirror systems, we choose the out-vacuum state as $\rho_0$ and the in-vacuum state as $\rho$. Then,
\begin{equation}
    \delta V=\begin{bmatrix}
        \delta a_1 &\delta a_3\\
        \delta a_3&\delta a_2
    \end{bmatrix},\quad\delta\nu=\frac{a_1\delta a_2+a_2\delta a_1-2a_3\delta a_3}{2\nu},
\end{equation}
where the variation of the covariance matrix is represented by the profile function $(Q(u),P(u))$ of the local mode:
\begin{subequations}
\begin{align}
    \delta a_1&=\int_{-\ell/2}^{\ell/2}
    du_1du_2\,Q(u_1)Q(u_2)[D_\text{in}(u_1+u_A,u_2+u_A)-D_{\text{out}}(u_1+u_A,u_2+u_A)],  \label{eq:da1}\\
      \delta a_2&=\int_{-\ell/2}^{\ell/2}
    du_1du_2\,P(u_1)P(u_2)[D_\text{in}(u_1+u_A,u_2+u_A)-D_{\text{out}}(u_1+u_A,u_2+u_A)], \label{eq:da2}\\
    \delta a_3&=\int_{-\ell/2}^{\ell/2}
    du_1du_2\,Q(u_1)P(u_2)[D_\text{in}(u_1+u_A,u_2+u_A)-D_{\text{out}}(u_1+u_A,u_2+u_A)].  \label{eq:da3}
\end{align}
\end{subequations}
Here, $D_{\text{out}}$ denotes the field correlation function evaluated in the out-vacuum, and $D_\text{in}$ is the corresponding correlation function defined with respect to the in-vacuum state.
Therefore, the variation of the entanglement entropy is rewritten as
\begin{equation}
    \delta S=\delta\langle H_A\rangle
    =\frac{\epsilon(\nu)}{4\nu}
    \left(a_2\delta a_1+a_1\delta a_2-2a_3\delta a_3\right).
\end{equation}
We introduce new integration variables $s,r$ in Eqs. \eqref{eq:da1},\eqref{eq:da2} and \eqref{eq:da3}:
\begin{equation}
  s=u_A+\frac{u_1+u_2}{2},\qquad r=u_1-u_2,\qquad -\ell\le r\le \ell,
\end{equation}
and approximate the variation of entanglement entropy under the assumption that the size $\ell$ of the local region is sufficiently small\footnote{We assume the size of the local region $\ell$ is larger than the UV cutoff $\ep$; thus, $\ep<\ell$.}. By expanding with respect to $r$, we obtain
\begin{align}
  D_\text{in}(u_1+u_A,u_2+u_A)&=-\frac{1}{2\pi}\frac{p'(s+\frac{r}{2})p'(s-\frac{r}{2})}
       {\left(p(s+\frac{r}{2})-p(s-\frac{r}{2})\right)^2} \notag \\
    &=
    -\frac{1}{2\pi}\left[\frac{1}{r^2}
    +\frac{1}{6}\frac{p'''}{p'}
    -\frac{1}{4}\left(\frac{p''}{p'}\right)^2\right]+O(\ell^2).
   \end{align}
Therefore, the difference in the correlation functions between in-vacuum and out-vacuum is   
\begin{align}
  D_\text{in}(u_1+u_A,u_2+u_A)-D_{\text{out}}(u_1+u_A,u_2+u_A)
    &=
    -\frac{1}{2\pi}
    \left[
      \frac{1}{r^2}
      -4\pi F(s)
      -\frac{1}{r^2}
    \right]+O(\ell^2)
    \notag\\
    &=
    2F\left(u_A+\frac{u_1+u_2}{2}\right)+O(\ell^2) \notag \\
    &=2F(u_A)+(u_1+u_2)F'(u_A)+O(\ell^2).
\end{align}
Consequently, $\del a_1,\del a_2,\del a_3$ are determined by the local value of the flux $F(u_A)$ at the leading order in $\ell$. Under this approximation, we arrive at
\begin{align}
	&\delta a_1\approx 2F(u_A)I_Q^2, \quad
	\delta a_3\approx 2F(u_A)I_QI_P,\quad
	\delta a_2\approx 2F(u_A)I_P^2,
\end{align}
where $I_Q,I_P$ are defined by
\begin{align}
    &I_Q=\int_{-\frac{\ell}{2}}^{\frac{\ell}{2}}\dd{u}Q(u)=\frac{4\ell}{\pi\sqrt{\pi}},\quad
    I_P=\int_{-\frac{\ell}{2}}^{\frac{\ell}{2}}\dd{u}P(u)=0,
\end{align}
and we have employed the profile functions~\eqref{eq:profile} in this step. 
Consequently, the variation of the entanglement entropy becomes
\begin{equation}
    \delta S\approx \frac{8\epsilon(\nu)}{\pi^3\nu}a_2\ell^2F(u_A).
\end{equation}
In this expression, because the coefficient of the flux is positive, the sign of $\delta S$ is completely controlled by the sign of the flux $F$. This clearly shows that the variation of entanglement entropy of the local region is directly linked to the change in the energy flux passing through it. 

\begin{figure}[thb]
\centering
\includegraphics[width=1\linewidth]{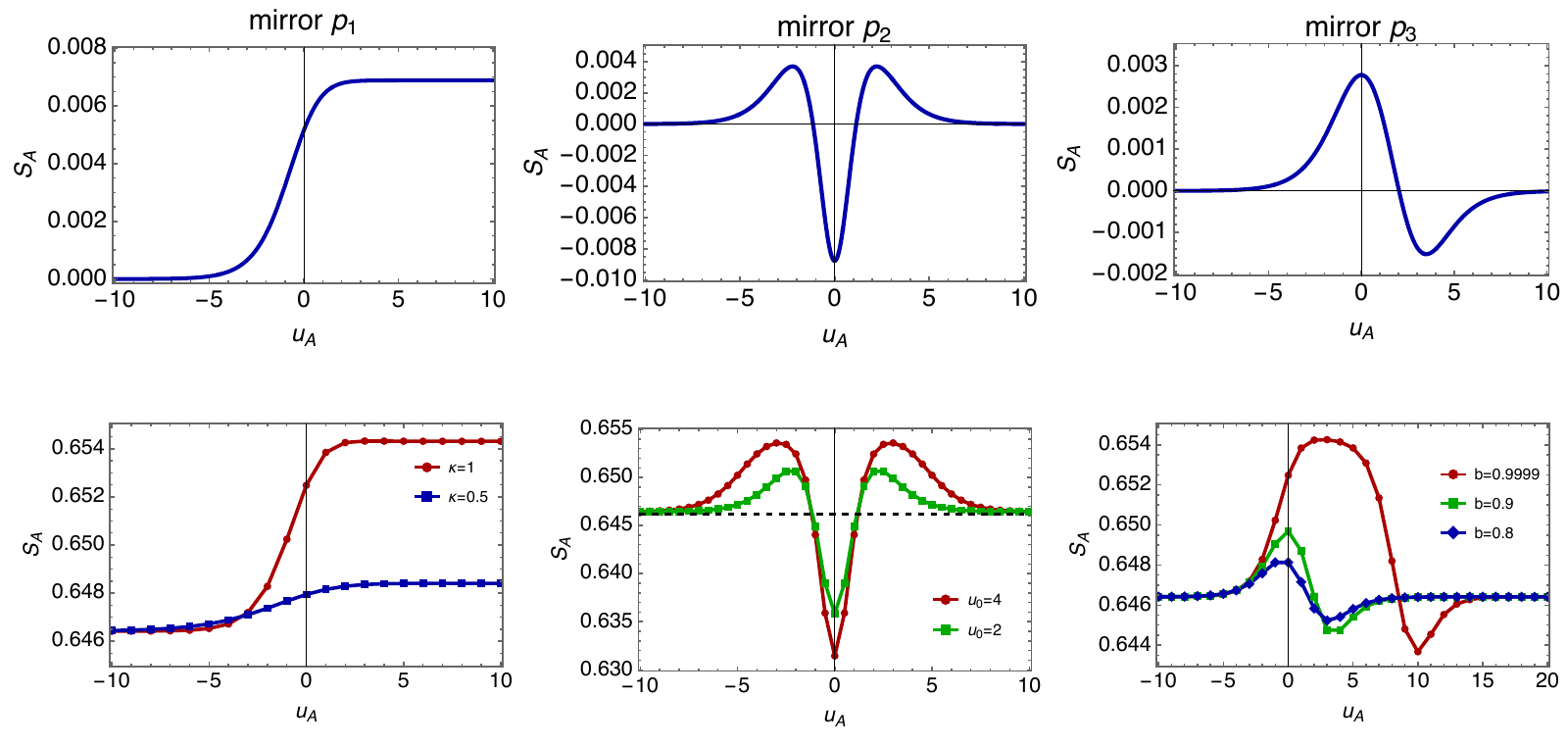}	
\caption{Upper panels: behavior of $S_A$ \eqref{eq:sA} for $\ell=1$. The parameters are set as follows: $\kappa=1$ for $p_1$, $\kappa=1, u_0=2$ for $p_2$, and $b=0.9$ for $p_3$. In the vacuum case, the value is shifted so that it vanishes, since the contribution from the out-vacuum state has been subtracted. For trajectories that generate negative energy flux, $S_A$ drops below its vacuum value during the interval in which the negative energy flux is emitted. Lower panels: $S_A$ obtained numerically from the symplectic eigenvalue of the local mode A. In this setting, the contribution from vacuum fluctuations is not subtracted.}
\label{fig:SA}
\end{figure}

To verify the aforementioned property, we plot the entanglement entropy associated with mode A.
The upper panels of Fig. \ref{fig:SA} present the behavior of $S_A$ in \eqref{eq:sA} computed without using the small-$\ell$ approximation, with $\ell$ set to 1. One can confirm that $S_A$ exhibits essentially the same behavior as $F(u)$, leading us to conclude that the relation \eqref{eq:1stlaw} is satisfied. The lower panels of Fig. \ref{fig:SA} depict $S_A$ obtained from the formula \eqref{eq:SA-nu} by a numerical evaluation of the symplectic eigenvalue of local mode A, rather than from the analytic formula \eqref{eq:S}. These numerical results show good agreement with the analytical prediction \eqref{eq:S}. We also find that the time dependence of $S_A$ closely tracks that of the energy flux $F(u)$ and the logarithmic negativity (Figs.~\ref{fig:trj1},\ref{fig:trj2},\ref{fig:trj3}). We have verified that this agreement holds at the level of the first law of entanglement entropy \cite{Faulkner2014}.

\subsubsection{Reduction and enhancement of bipartite entanglement}
We observe that the passage of the negative energy flux reduces $S_A$, which represents a reduction in the entanglement between the local mode A and its complement. For a Gaussian pure system composed of a bipartite system AB and its complement C, the following monogamy inequality holds for entanglement measures $E(A:B)$ and $E(AB:C)$ \cite{Camalet2018,Nambu2023b}
\begin{equation}
E(A:B)+E(AB:C)\leq E_\text{max},
\end{equation}
where $E_\text{max}$ is the maximum of $E(A:B)$. This inequality represents a trade-off relation between the internal correlation $E(A:B)$ and the external correlation $E(AB:C)$.
Therefore, by analyzing the monogamy relation between the local modes A and B and their complementary subsystems \cite{Nambu2023b}, the decrease of $S_A$ (and $S_B$) can account for the potential enhancement of quantum correlations between A and B, which consequently results in a larger negativity of the bipartite state AB. This accounts for the increase in negativity when a negative energy flux is present because the energy flux of the radiation is exactly the entanglement current or information flux, and it controls the behavior of the internal entanglement of the bipartite detector mode AB.

This finding indicates that energy flux encompasses both the transfer of energy and the transmission of information.
In the region carrying positive energy flux, the radiation degrees of freedom become entangled with other degrees of freedom located outside the radiation zone; consequently, the internal entanglement between A and B is diminished. In contrast, in the region with negative energy flux, the internal correlation between A and B is restored, leading to an increase in the negativity. Thus, one may regard the negative energy flux as an “information return channel.”

\subsubsection{Relation to Partner Modes}
To further support our interpretation that the negative energy flux serves as the channel through which information returns, we analyze the previously discussed entanglement property from the viewpoint of the entangled partner.
To discuss how information can be extracted from the radiation emitted by the mirror, we analyze the profile of the partner mode P that purifies the detector mode A. Modes A and P together constitute a pure two-mode squeezed state.  
For given mode A, characterized by the profile functions $(Q_A(u), P_A(u))$, the corresponding profile of the partner mode P is given by \cite{Tomitsuka2019} (see Appendix \ref{sec:appendix} for a review of the partner formula)\footnote{This partner formula essentially implements the Hilbert transform of the detector profile functions $(Q_A(u),P_A(u))$. Under the Hilbert transform, a function with compact support is mapped to one whose support is no longer compact, highlighting the inherently non-local character of entanglement.}
\begin{subequations}
\begin{align}
\begin{bmatrix}
Q_P(u) \\ P_P(u)
\end{bmatrix}&=\frac{\nu}{\sqrt{\nu^2-1}}
\begin{bmatrix}
Q_A(u) \\ -P_A(u)
\end{bmatrix}
+\frac{1}{\sqrt{\nu^2-1}}\int_{-\infty}^\infty du_1\pa_{u_1}p(u_1)\Delta[p(u)-p(u_1)]
\begin{bmatrix}
P_A(u_1) \\ Q_A(u_1)
\end{bmatrix},
\label{eq:partner-formula}
\\
&\Delta[x-y]:=\frac{1}{2\pi}\left(\frac{1}{x-y-i\ep}+\frac{1}{x-y+i\ep}\right)
=\frac{1}{\pi}\text{P.V.}\left(\frac{1}{x-y}\right),\quad\ep>0,
\end{align}
\end{subequations}
where $\nu$ is the symplectic eigenvalue of the detector mode A given by \eqref{eq:nuA}.
The form of the partner profiles encodes how  another detector can extract information about the entanglement partners of Hawking radiation. For an entanglement harvesting problem with two detector modes A and B, one can analyze the detectable entanglement of the radiation  by examining the overlap between the partner profile functions and the profile functions of detector B. One can define a discriminant that measures how much the partner mode P overlaps with the detector mode B \cite{Osawa2026}.

\begin{figure}[H]
\centering
\includegraphics[width=0.9\linewidth]{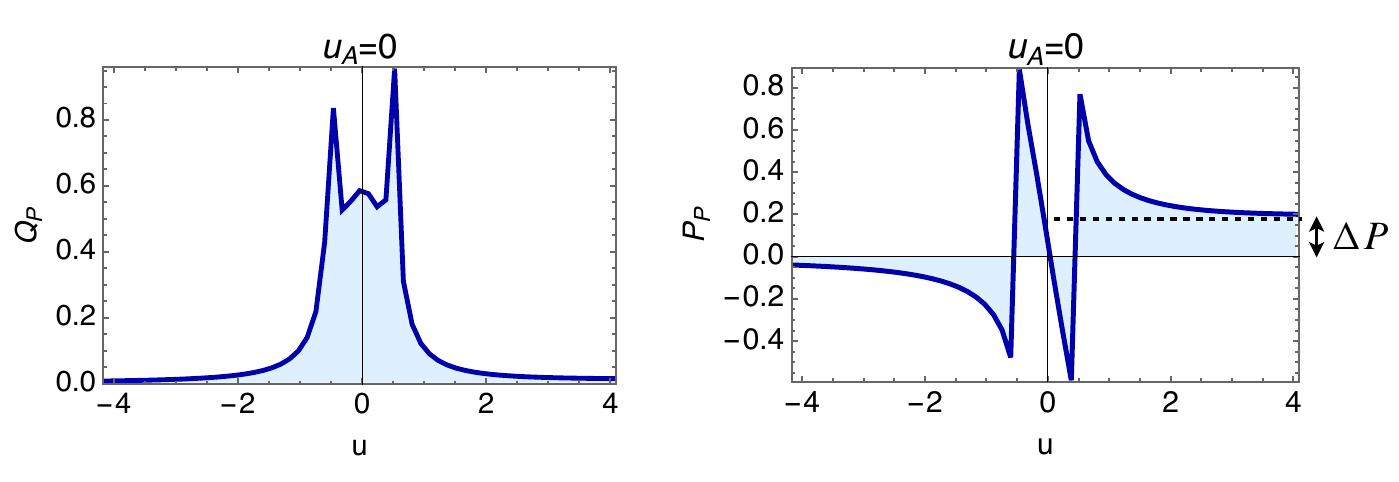}	
\caption{The typical shape of the partner profile functions is illustrated for mirror $p_1$ with $u_A = 0$. Although $Q_P(u)$ is symmetric around $u = u_A$, $P_P(u)$ displays a non-decaying, long-range offset in the domain $u_A< u$. We introduce the offset of $P_P(u)$ from the $u$-axis at $u = u_A + \Delta u$, denoted by $\Delta P$, whose magnitude is determined by $u_A$. In our numerical analysis, we set $\Delta u = 4.13$. This offset characterizes the degree of non-local correlations between detector mode A and its partner mode. A larger value of $\Delta P$ indicates stronger long-range quantum correlations, which in turn increases the likelihood that the harvesting protocol involving detectors A and B will fail. }
\label{fig:partner-profile}
\end{figure}

\begin{figure}[bht]
\centering
\includegraphics[width=0.9\linewidth]{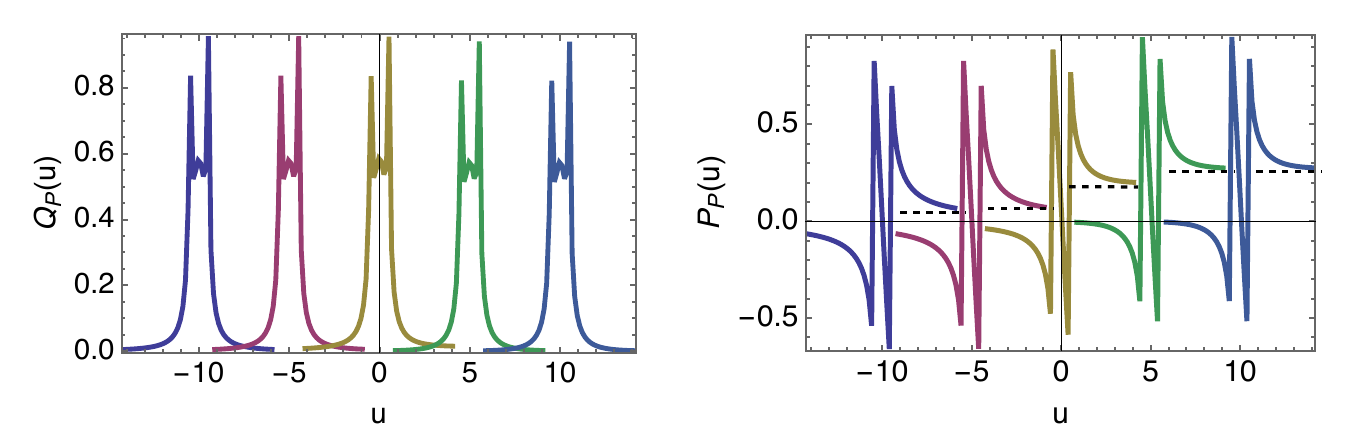}	
\caption{The partner profiles corresponding to mirror $p_1$ are shown for several choices of $u_A$ (left: $Q_P(u)$, right: $P_P(u)$). Detector A is located at $u_A = -10, -5, 0, 5, 10$. While the shape of $Q_P(u)$ remains unchanged over time, the offset $\Delta P$ of $P_P(u)$ increases as time progresses, revealing a non-local correlation between detector mode A and its associated partner mode.}
\label{fig:p1-profile}
\end{figure}

Figure \ref{fig:partner-profile} shows that the typical shape of the partner profile functions for the case of the mirror $p_1$  with $u_A=0$. For both $Q_P(u)$ and $P_P(u)$, the amplitudes diverge at the edges of region A, $u = \pm \ell/2$, indicating that the entanglement shared between A and P is concentrated near the boundaries of the region. While $Q_P(u)$ is symmetric about $u=u_A$, $P_P(u)$ exhibits a non-decaying long-range offset in the domain $u_A<u$. We introduce the amount of offset of $P_P(u)$ from the $u$-axis at a pivot point $u=u_A+\Delta u$, denoted by $\Delta P:=P_P(u_A+\Delta u)$. It can be shown  that  $\Delta P\propto \ell\,\kappa(u_A)$ with $\kappa(u_A)=-p''(u_A)/p'(u_A)$, whose magnitude depends on $u_A$ (see Appendix \ref{sec:appendix}). In our numerical calculations, we choose $\Delta u=4.13$. This offset $\Delta P $ implies the presence of non-local correlations between detector mode A and its partner mode (see Appendix \ref{sec:appendix} for a more detailed discussion). When $\Delta P$ is larger, stronger long-range quantum correlations between A and P are present, and the harvesting protocol using detectors A and B is more likely to fail because ``density" of the partner mode is diluted \cite{Osawa2026}. 
Figure \ref{fig:p1-profile} illustrates the partner profiles of mirror $p_1$ for different values of $u_A$ (left: $Q_P(u)$, right: $P_P(u)$). Detector A is placed at $u_A = -10, -5, 0, 5, 10$. While the shape of $Q_P(u)$ remains unchanged in time, the offset $\Delta P$ of $P_P(u)$ increases as time progresses, signaling a non-local correlation between detector mode A and its partner mode.

\begin{figure}[bht]
\centering
\includegraphics[width=0.9\linewidth]{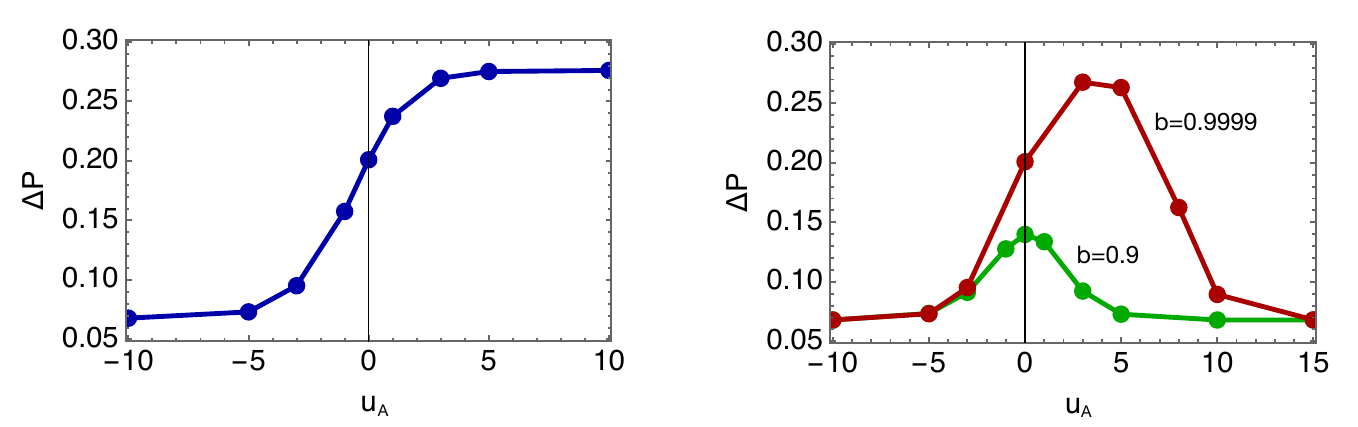}	
\caption{ The behavior of $\Delta P$ is shown for mirror $p_1$ (left panel) and mirror $p_3$ (right panel). For mirror $p_1$ (the eternal mirror), $\Delta P$ approaches a finite, non-zero constant, implying that non-local correlations persist and the information shared between detector A and its partner cannot be retrieved. In contrast, for mirror $p_3$, $\Delta P$ initially grows, then declines, and ultimately returns to its initial value. The time at which $\Delta P$ reaches its maximum coincides with the onset of the emission of negative energy flux. In this mirror model, the information shared between detector mode A and its partner mode is restored as the negative energy is emitted.}
\label{fig:delP}
\end{figure}

Figure \ref{fig:delP} shows how $\Delta P$ evolves for mirror $p_1$ (left panel) and mirror $p_3$ (right panel). For mirror $p_1$ (the eternal mirror), $\Delta P$ asymptotically approaches a finite, non-zero value, indicating that non-local correlations survive and that the information shared between detector A and its partner cannot be retrieved on $\mathscr{I}^+$. In this case, the information is effectively lost. 
By contrast, for mirror $p_3$, $\Delta P$ initially increases, then decreases, and eventually returns to its original value. The moment at which $\Delta P$ attains its maximum coincides with the onset of negative energy flux emission. In this setup, the information shared between detector mode A and its partner mode is completely restored once the negative energy flux appears. Altogether, these plots reproduce the characteristic Page-curve behavior and demonstrate how information is redistributed in the moving mirror models. 

Thus, through analyzing the behavior of the partner profile, we have confirmed our claim that the negative energy flux serves as an “information return channel” in moving mirror systems.

\section{Summary and conclusion}

 We studied the entanglement properties of analog Hawking radiation using a moving mirror setup. Implementing an entanglement harvesting scheme with a pair of detector modes AB placed at $\mathscr{I}^+$, we examined how the negativity of the bipartite state AB responds to the quantum radiation produced by the mirror. The sign of the energy flux emitted by the mirror is governed by the sign of the time variation of the mirror's acceleration. We found that when the mirror emits negative energy flux, the observed negativity increases, whereas for positive energy flux, the negativity decreases. We interpreted the behavior of negativity using the first law of entanglement. Enforcing information conservation (unitarity), we deduced that the production of negative energy is inherently associated with the retrieval of the corresponding partner particles.

To further support our interpretation that the negative energy flux functions as the information return channel, we employed the partner formula and tested the reconstruction of the partner mode by analyzing the profile of the partner mode that purifies detector mode A. We observed that the offset of the partner profile \(P_P(u)\) determines the spatial extent of the partner mode and measures how much of the correlation lies outside the reach of the AB detectors. As a result, we find that this offset exhibits a Page-curve-like behavior, signaling information retrieval for asymptotically timelike mirrors. In the setting of evaporating black holes, we therefore anticipate that the partner formula will serve as a valuable method for examining the behavior of non-local quantum field correlations, as demonstrated in this work, and this will be the central topic of our upcoming study on the subject.

\vspace{1ex}
After we uploaded our paper to arXiv, we became aware of the work \cite{Dubey_2026}. In that paper, the authors study memory (non-Markovian) effects in entanglement harvesting using qubit detectors. As one application, they analyze a moving mirror setup and find that the amount of entanglement that can be harvested increases when the mirror emits negative energy flux, although they do not discuss the underlying mechanism of this phenomenon in detail.


\begin{acknowledgments}
We would like to thank Yuki Osawa for his contributions during the initial stage of this project, and Kotaro Kondo in National Institutes for Quantum Science and Technology (QST) for his insightful contributions and discussions related to this project. We would also like to thank Masahiro Hotta for his comments on this subject. RY was supported by JSPS KAKENHI (Grant No.~26KJ1356) and YN was partly supported by JSPS KAKENHI (Grant No.~JP23K25871). This work was supported by JST FOREST Program, Grant Number JPMJFR2363.
\end{acknowledgments}

\bibliography{./partner.bib}

\appendix
\section{Partner formula}\label{sec:appendix}
\subsection{Derivation}
Consider a pure two-mode Gaussian state (i.e., a system of two harmonic oscillators) characterized by two independent annihilation operators, $\hat a_1$ and $\hat a_2$. We take the modes associated with $\hat a_1$ and $\hat a_2$ to be in their respective vacuum states, which satisfy the conditions $\hat a_1\ket{0_1}=\hat a_2\ket{0_2}=0$. The combined vacuum state of the system is then $\ket{\psi}:=\ket{0_10_2}$. We now define a new mode A as
\begin{equation}
 \hat a_A=\alpha\,\hat a_1+\beta\,\hat a{}_2{}^{\!\dag},\quad |\alpha|^2-|\beta|^2=1,\quad [\hat a_1,\hat a_2]=[\hat a_1,\hat a_2{}^{\!\dag}\,]=0,\quad [\hat a_A,\hat a_A{}^{\!\dag}\,]=1,
 \label{eq:eqA}
\end{equation}
where, for convenience, we assume the coefficients $\alpha,\beta$ are real numbers and ignore any possible phase factors. The mode defined by the annihilation operator below then provides a purification of mode A:
\begin{equation}
 \hat a_P=\beta\,\hat a_1{}^{\!\dag}+\alpha\,\hat a_2,\quad
 [\hat a_A,\hat a_P]=[\hat a_A,\hat a_P{}^{\!\dag}]=0,\quad
 [\hat a_P,\hat a_P{}^{\!\dag}]=1.
 \label{eq:eqP}
\end{equation}
This mode is identified as the partner of mode A. Indeed, we will show that mode AP constitutes a true two-mode squeezed state. The designation of this mode as the partner of A arises because the covariance matrix of AP has the same structure as that of a pure two-mode squeezed state (see \eqref{eq:two-squeeze}). Consequently, mode P provides a purification of mode A. By inverting the relations \eqref{eq:eqA} and \eqref{eq:eqP}, we find
\begin{equation}
  \hat a_1=\alpha\,\hat a_A-\beta\,\hat a_P{}^{\!\dag},\quad \hat
  a_2^\dag=-\beta\,\hat a_A+\alpha\,\hat a_P{}^{\!\dag}.
\end{equation}
We can verify that the two modes, A and P, form a pure state as follows: We introduce the canonical operators for modes A, P, 1, and 2:
\begin{equation}
  \hat Q_{j}=\frac{\hat a_{j}+\hat a_{j}{}^{\!\dag}}{\sqrt{2}},\quad
  \hat P_{j}=\frac{\hat a_{j}-\hat
    a_{j}{}^{\!\dag}}{i\sqrt{2}},\quad j=A,P,1,2. 
\end{equation}
Then, the expressions for mode A and mode P are given by
\begin{subequations}
\begin{align}
  &\hat Q_A=\alpha\,\hat Q_1+\beta\,\hat Q_2,\quad \hat P_A=\alpha\,\hat
    P_1-\beta\,\hat P_2, \label{eq:QA}\\
  &\hat Q_P=\beta\,\hat Q_1+\alpha\,\hat Q_2,\quad \hat P_P=-\beta\,\hat
    P_1+\alpha\,\hat P_2. \label{eq:QB}
\end{align}
\end{subequations}
We now adopt a vector notation for these canonical operators:
\begin{equation}
  \hat{\bm{r}}:=(\hat Q_1,\hat P_1,\hat Q_2,\hat P_2)^T,\quad
  \hat{\bm{\xi}}_A:=(\hat Q_A,\hat P_A)^T,\quad
  \hat{\bm{\xi}}_P:=(\hat Q_P,\hat P_P)^T.
\end{equation}
The commutator and the covariance matrix associated with $\hat{\bm{r}}$ are\footnote{In component form, one has $[\hat r^a,\hat r^b]=i\,J^{ab}$.}
\begin{equation}
[\hat{\bm{r}},\hat{\bm{r}}^T]=i\,\bm{J},\quad  \bm{M}_{12}:=\expval{\{\hat{\bm{r}},\hat{\bm{r}}^T\}}= \bm{I}_4,
\end{equation}
where
\begin{equation}
  \bm{J}:=\bigoplus_{i=1}^2\bm{\Omega}=
  \begin{bmatrix}
    0&1&0&0 \\-1&0&0&0 \\ 0&0&0&1 \\ 0&0&-1&0
  \end{bmatrix}
  ,\quad\bm{\Omega}:=
  \begin{bmatrix}
    0&1\\-1&0
  \end{bmatrix}.
\end{equation}
The covariance matrix $\bm{M}_\text{AP}$ for the bipartite mode  $\hat{\bm{\xi}}_{AP}=(\hat
Q_A,\hat P_A,\hat Q_P,\hat P_P)^T$ is
\begin{align}
  \bm{M}_\text{AP}=\expval{\{\hat{\bm{\xi}}_{AP},\hat{\bm{\xi}}_{AP}^T\}}_\psi&=
  \begin{bmatrix}
    \alpha^2+\beta^2&0&2\alpha\beta&0\\
    0&\alpha^2+\beta^2&0&-2\alpha\beta\\
    2\alpha\beta&0&\alpha^2+\beta^2&0\\
    0&-2\alpha\beta&0&\alpha^2+\beta^2
  \end{bmatrix}
=
    \begin{bmatrix}
      \cosh 2r&0&\sinh 2r&0\\
    0&\cosh 2r&0&-\sinh 2r\\
    \sinh 2r&0&\cosh 2r&0\\
    0&-\sinh 2r&0&\cosh 2r
  \end{bmatrix}, \label{eq:two-squeeze}
\end{align}
Here, $\expval{\cdot}_\psi$ denotes the expectation value with respect to the vacuum state $\ket{\psi}$, and the parameter $r$ is defined through $\alpha=\cosh r$ and $\beta=\sinh r$. For simplicity, we will omit the subscript and write $\expval{\cdot}$ instead of $\expval{\cdot}_\psi$ from now on. The quantity $\nu := \cosh 2r$ is the symplectic eigenvalue associated with modes A and P. The covariance matrix describes a pure two-mode squeezed state. Consequently, mode P provides a purification of the original mode A and is referred to as its partner. The partner formula specifies mode P in terms of mode A. To derive the explicit form of this partner formula, we introduce the following linear map $f_\psi$ acting on the pure-state quadrature vector
$\hat{\bm{r}} = (\hat Q_{1}, \hat P_1, \hat Q_{2}, \hat P_2)^T$:
\begin{equation}
  f_\psi(\hat Q_j)=\hat P_j,\quad f_{\psi}(\hat P_j)=-\hat Q_j,\quad j=1,2.
\end{equation}
In a more compact notation, the map can be written as
\begin{equation}
  f_\psi(\hat{\bm{r}})=\bm{J}\hat{\bm{r}}.
\end{equation}
This map defines a “complex structure”\footnote{This means that the algebraic relation $\bm{J}^2=-\bm{I}$ is satisfied.}
on the total pure system
\cite{Hackl2019}. More precisely, for a pure state, the annihilation operator of mode $j=1,2$ can be defined as
$\hat a_j:=(\hat Q_j+if_\psi(\hat Q_j))/\sqrt{2}$. Using Eqs. \eqref{eq:QA} and \eqref{eq:QB}, the resulting action of this map on $\hat Q_A$ and $\hat P_A$ is
\begin{equation}
  f_\psi(\hat Q_A)=\alpha\,\hat P_1+\beta\,\hat P_2,\quad
  f_\psi(\hat P_A)=-\alpha\,\hat Q_1+\beta\,\hat Q_2.
  \label{eq:ff2}
\end{equation}
For pure state operators $\hat a_j,\hat a_j{}^{\!\dag}$,
\begin{equation}
  f_\psi(\hat a_j)=-i\,\hat a_j,\quad f_\psi(\hat a_j{}^{\!\dag})=i\,\hat
  a_j{}^{\!\dag},\quad j=1,2.
\end{equation}

 For arbitrary Hermitian
operators $\hat O_1, \hat O_2$ defined by linear combinations of the pure
state basis $(\hat Q_1,\hat P_1,\hat Q_2,\hat P_2)$, it can be shown
that this map has the following properties:\cite{Yamaguchi2020}
\begin{subequations}
\begin{align}
  [\hat O_1,f_\psi(\hat O_2)]&=i\expval{\{\hat O_1,\hat O_2\}}, \quad [f_\psi(\hat O_1),f_\psi(\hat O_2)]=[\hat O_1,\hat O_2], \label{eq:f1}\\
  \expval{\{\hat O_1,f_\psi(\hat O_2)\}}&=i[\hat O_1,\hat O_2],\quad
  \expval{\{f_\psi(\hat O_1),f_\psi(\hat O_2)\}}
  =\expval{\{\hat O_2,\hat O_2\}},  \label{eq:f4} \\
  f_\psi^2(\hat O)&=-\hat O. \label{eq:f5}
\end{align}
\end{subequations}
Using \eqref{eq:QA} and \eqref{eq:ff2}, the operators $(\hat Q_{1}, \hat
P_{1},\hat Q_2,\hat P_2)$ are expressed using $\hat Q_A, \hat P_A$ as
\begin{subequations}
\begin{align}
  &\hat Q_1=\frac{1}{2\alpha}(\hat Q_A-f_\psi(\hat P_A)),\quad
    \hat Q_2=\frac{1}{2\beta}(\hat Q_A+f_\psi(\hat P_A)),\\
  &\hat P_1=\frac{1}{2\alpha}(f_\psi(\hat Q_A)+\hat P_A),\quad
    \hat P_2=\frac{1}{2\beta}(f_\psi(\hat Q_A)-\hat P_A).
\end{align}
\end{subequations}
In terms of the annihilation operator,
\begin{equation}
  \hat a_1=\frac{1}{2\alpha}\left(\hat a_A+if_\psi(\hat a_A)\right),\quad
  \hat a_2=\frac{1}{2\beta}\left(\hat a_A{}^{\!\dag}+if_\psi(\hat
    a_A{}^{\!\dag})\right).
  \label{eq:pure-mode}
\end{equation}
Substituting these relations into Eq. \eqref{eq:QB},
\begin{subequations}
\begin{align}
  &\hat
    Q_P=\frac{1}{2}\left(\frac{\alpha}{\beta}+\frac{\beta}{\alpha}\right)\hat
    Q_A+\frac{1}{2}\left(\frac{\alpha}{\beta}-\frac{\beta}{\alpha}\right)f_\psi(\hat
    P_A), \\
   &\hat
    P_P=-\frac{1}{2}\left(\frac{\alpha}{\beta}+\frac{\beta}{\alpha}\right)\hat
    P_A+\frac{1}{2}\left(\frac{\alpha}{\beta}-\frac{\beta}{\alpha}\right)f_\psi(\hat
    Q_A).
\end{align}
\end{subequations}
Using  the symplectic eigenvalue $\nu$ of mode A, $\alpha^2=(\nu+1)/2,
\beta^2=(\nu-1)/2$ and  we obtain a relation connecting mode A and mode P:
\begin{equation}
  \hat Q_\text{P}=\frac{\nu}{\sqrt{\nu^2-1}}\hat
  Q_\text{A}+\frac{f_\psi(\hat
    P_\text{A})}{\sqrt{\nu^2-1}},
  \quad
    \hat P_\text{P}=-\frac{\nu}{\sqrt{\nu^2-1}}\hat P_\text{A}+\frac{f_\psi(\hat
      Q_\text{A})}{\sqrt{\nu^2-1}}.
    \label{eq:partner-f}
\end{equation}
This is the partner formula expressed in a basis-independent way, which captures the structure of the two-mode squeezed state (the partner structure). The map $f_\psi$ determines how a local mode is embedded into a globally pure state, and its form is dictated by the specific details of the total pure system. In terms of the annihilation operators for each mode, the partner formula can be written in the following simple form:
\begin{equation}
  \hat a_P=\frac{\nu}{\sqrt{\nu^2-1}}\,\hat
  a_A{}^{\!\dag}+\frac{i}{\sqrt{\nu^2-1}}f_\psi(\hat a_A{}^{\!\dag}).
  \label{eq:partner2}
\end{equation}

As an example, we examine the partner formula for a  chiral scalar field in (1+1)-dimensional Minkowski spacetime with the mirror boundary condition. Using the in-mode basis on $\mathscr{I}^+$,  the field operators are given by
\begin{subequations}
\begin{align}
 &\hat\phi_\text{in}(v)=\int_0^\infty\frac{d\omega}{\sqrt{4\pi\omega}}\left(\hat a_\omega e^{-i\omega v}+\hat a_\omega{}^{\!\dag} e^{i\omega v}\right),\quad v=t+x, \\
 &\hat\Pi_\text{in}(v)=\partial_v\hat\phi_\text{in}=-i\int_0^\infty d\omega\sqrt{\frac{\omega}{4\pi}}\left(\hat a_\omega e^{-i\omega v}-\hat a_\omega{}^{\!\dag}\, e^{i\omega v}\right).
\end{align}
\end{subequations}
where the in-vacuum is defined by $\hat a_\omega\ket{0_\text{in}}=0$. The linear map $f_\psi$ acts as $f_\psi(\hat a_\omega)=-i\hat a_\omega, f_\psi(\hat a_\omega{}^{\!\dag})=i\hat a_\omega{}^{\!\dag}$. Actually, for an arbitrary Hermitian operator $\hat O$ defined by the profile function $O^\text{in}(v)$ as $\hat O=\int_{-\infty}^\infty dv\, O^\text{in}(v)\hat\Pi_\text{in}(v)$,
we can prove the properties \eqref{eq:f1}-\eqref{eq:f5}. Using this map $f_\psi$, for local canonical operators $(\hat Q_A,\hat P_A)$, it is possible to show \cite{Tomitsuka2019}
\begin{subequations}
\begin{align}
 &f_\psi\circ\begin{bmatrix} \hat Q_A \\ \hat P_A\end{bmatrix}=\int_{-\infty}^\infty dv_1dv_2\hat\Pi_\text{in}(v_2)\Delta(v_2-v_1)\begin{bmatrix}Q_A^\text{in}(v_1) \\ P_A^\text{in}(v_1)\end{bmatrix},\\
 &\Delta(v_2-v_1):=2\expval{\{\hat\phi(v_1),\hat\Pi(v_2)\}}=\frac{1}{\pi}\text{P.V.}\left(\frac{1}{v_2-v_1}\right).
\end{align}
\end{subequations}
From the above equations, we derive the partner formula for the mode profile functions:
\begin{align}
\begin{bmatrix}
Q_P^\text{in}(v) \\ P_P^\text{in}(v)
\end{bmatrix}&=\frac{\nu}{\sqrt{\nu^2-1}}
\begin{bmatrix}
Q_A^\text{in}(v) \\ -P_A^\text{in}(v)
\end{bmatrix}
+\frac{1}{\sqrt{\nu^2-1}}\int_{-\infty}^\infty dv_1\Delta[v-v_1]
\begin{bmatrix}
P_A^\text{in}(v_1) \\ Q_A^\text{in}(v_1)
\end{bmatrix}.
\label{eq:partner-formula0}
\end{align}
To obtain the mode profiles on $\mathscr{I}^+$, we consider the relation between the profile functions on $\mathscr{I}^-$ (in-mode) and $\mathscr{I}^+$ (out-mode):
\begin{equation}
 Q_j^\text{out}(u)=-Q_j^\text{in}(p(u)),\quad P_j^\text{out}(u)=-P_j^\text{in}(p(u)),\quad j=A,P,
\end{equation}
which arise from the boundary condition at the mirror. Next, by substituting the integration variable $v_1$ with $u_1$ using the relation $v_1 = p(u_1)$ in \eqref{eq:partner-formula0}, we arrive at \eqref{eq:partner-formula}.
\subsection{Profile of partner mode}
To analyze the characteristics of the partner mode predicted by the partner formula \eqref{eq:partner-formula}, we introduce local mode A, defined via the following simple profile function, which allows us to determine the partner mode profile analytically:
\begin{align}
&Q_A(u)=\sqrt{\frac{2}{\ell}}\,\chi_{[-\ell/2,\ell/2]}(u-u_A),\quad P_A(u)=\sqrt{\frac{2}{\ell}}\,(u-u_A)\,\chi_{[-\ell/2,\ell/2]}(u-u_A),
\end{align}
where $\chi_{[a,b]}(u)$ is the box function defined by
\begin{equation}
 \chi_{[a,b]}(u)=
 \begin{cases}
 1\quad& u\in[a,b] \\
 0\quad& u\notin[a,b]
 \end{cases}.
\end{equation}
Then, in the partner formula \eqref{eq:partner-formula}, at a pivot point $u=u_A+\Delta u$ with  $\ell<\Delta u$ and $\ell\ll p/p'$, the integrals   are evaluated as follows:
\begin{align}
\begin{bmatrix}
I_1(u)\\ I_2(u)
\end{bmatrix}
&=
\int_{u_A-\ell/2}^{u_A+\ell/2} du_1\,\pa_{u_1}p(u_1)\frac{1}{p(u)-p(u_1)}
\begin{bmatrix}
P_A(u_1) \\ Q_A(u_1)
\end{bmatrix} \notag \\
&=\int_{-\ell/2}^{\ell/2} dx\, p'(u_A+x)\frac{1}{p(u)-p(u_A+x)} 
\begin{bmatrix}
x \\ 1
\end{bmatrix} \notag \\
&\approx
\begin{bmatrix}
\dfrac{\ell^3}{12(\Delta u)^2} \\
\dfrac{\ell}{\Delta u}-\ell\dfrac{p''(u_A)}{2p'(u_A)}
\end{bmatrix}
\sim
\begin{bmatrix}
0 \\
\ell\,\kappa(u_A)/2
\end{bmatrix}\quad\text{for}\quad \ell/\Delta u\ll 1,
\end{align}
where $\kappa(u)=-p''(u)/p'(u)$ is the peeling factor \cite{Bianchi2014a}, which, in the case of the eternal mirror $p_1$, matches the surface gravity associated with the Rindler horizon.
 Consequently, the partner profile $P_P(u)$ develops a non-vanishing offset in the domain $u_A<u$, whose magnitude is determined by the peeling factor and it reduces to the surface gravity of the Rindler horizon for the eternal mirror $p_1$. The presence of a nonzero offset in the profile shows that the mode is infinitely spread out; as a result, in the usual sense, there is no corresponding partner mode on $\mathscr{I}^+$ in this case (information loss) because the partner mode cannot be realized with a profile function with compact support. Nonetheless, the way the partner profile behaves reflects this absence of a genuine partner mode through a singular feature in its profile.

Even in the case of asymptotically time-like mirrors, the temporal offset of the profile $P_p(u)$ related to the peeling factor $\kappa(u_A)$  appears in the domain $u_A<u$,  and this pattern signals a temporal establishment of the correlations associated with the partner mode.

\section{Derivation of $\Tr \left[\rho_0\,\delta\!\log \rho\right]=0$}\label{sec:appendixB}
Here, we will show $\Tr \left[\rho_0\,\delta\!\log \rho\right]=0$.
First, we diagonalize the initial density matrix as
\begin{equation}
    \rho_0=\sum_np_n\ket{n}\bra{n},
\end{equation}
and also introduce the eigenstates and eigenvalues of the $\rho$ as
\begin{equation}
        \rho\ket{n'}=p'_n\ket{n'},\quad
        p'_n=p_n+\delta p_n,\quad
        \ket{n'}=\ket{n}+\ket{\delta n}.
\end{equation}
We assume that the spectrum $\{p_n\}$ is not degenerate; however, as will be explained below, only the diagonal components of $\delta(\log \rho)$ contribute to the final result, so the result would remain the same even if it were degenerate.
We need to evaluate the deviation from $\ket{n},\ p_n$, and this is carried out using the same type of calculation as in standard perturbation theory:
\begin{equation}
    (\rho_0+\delta \rho)(\ket{n}+\ket{\delta n})=(p_n+ \delta p_n)(\ket{n}+\ket{\delta n}).
\end{equation}
To first order in the perturbation,we obtain
\begin{equation}
    p_m\langle m |\delta n\rangle+X_{mn}=p_n\langle m|\delta n\rangle+\delta p_n\delta_{mn},
\end{equation}
where $X_{mn}=\langle m|\delta \rho |n\rangle$. Focusing on the diagonal components, we find
\begin{equation}
    \delta p_n=X_{nn}.
\end{equation}
On the other hand, the off-diagonal components become
\begin{equation}
    \langle m|\delta n\rangle =\frac{X_{mn}}{p_n-p_m}.
\end{equation}
Using these equations allows us to compute $\Tr\left[\rho_0\,\delta\log \rho\right]$. To this end, we examine the spectral decomposition of $\log \rho$:
\begin{equation}
    \log \rho=\sum_n\log p'_n \ket{n'}\bra{n'}
\end{equation}
To the first order, using the relations
\begin{equation}
    \begin{split}
        \log p'_n&=\log p_n + \frac{\delta p_n}{p_n},\\
        \ket{n'}\bra{n'}&=\ket{n}\bra{n}+(\ket{\delta n}\bra{n}+\ket{n}\bra{\delta n}),
    \end{split}
\end{equation}
the variation of $\log \rho$ becomes
\begin{equation}
    \delta (\log \rho)=\sum_n \left[\frac{\delta p_n}{p_n}\ket{n}\bra{n}+\log p_n  (\ket{\delta n}\bra{n}+\ket{n}\bra{\delta n}) \right].
\end{equation}
First, let us consider the diagonal components. Then, the second term becomes zero because of the normalization condition of the trace of the density matrix. Therefore, we obtain
\begin{equation}
    [\delta (\log \rho)]_{nn}=\frac{X_{nn}}{p_n}.
\end{equation}
For the off-diagonal components, the first term is zero, and thus we get
\begin{equation}
\begin{split}
    [\delta(\log \rho)]_{mn}&=\log p_n \langle m|\delta n\rangle +\log p_m \langle \delta m|n\rangle\\
    &=\log p_n\frac{X_{mn}}{p_n-p_m}+\log p_m \frac{X_{mn}}{p_m-p_n}\\
    &=\frac{\log p_m-\log p_n}{p_m-p_n}X_{mn}.
    \end{split}
\end{equation}
From this equation, we obtain
\begin{equation}
    \begin{split}
        \Tr \left[\rho_0\,\delta\!\log \rho\right]&=\sum_n\bra{n}\rho_0\delta(\log \rho)\ket{n}=\sum_np_n\bra{n}\delta(\log \rho)\ket{n}\\
        &=\sum_np_n\frac{X_{nn}}{p_n}=\sum_nX_{nn}\\
        &=\Tr [\delta \rho]=0.
    \end{split}
\end{equation}
Although this calculation assumes that $p_n$ is non-degenerate, even in the presence of degeneracy the diagonal elements of $\delta\log\rho$ are unaffected, and therefore the above results still hold.

\end{document}